\documentclass[pra]{revtex4}
\usepackage{graphicx}
\usepackage{epsfig}

\newcommand{\beq}{\begin{equation}}
\newcommand{\eeq}{\end{equation}} 
\newcommand{\beqa}{\begin{eqnarray}}
\newcommand{\eeqa}{\end{eqnarray}}
\newcommand{\ba}{\begin{array}}
\newcommand{\ea}{\end{array}}

\begin{document}

\title{DC and AC Josephson effects with superfluid Fermi atoms \\ 
across a Feshbach resonance} 
\author{L. Salasnich$^{1,2}$, F. Ancilotto$^2$, 
N. Manini$^3$, and F. Toigo$^2$} 
\affiliation{$^1$CNR-INFM and CNISM, Unit\`a di Padova, \\ 
Via Marzolo 8, 35122 Padova, Italy \\
$^2$Dipartimento di Fisica ``Galileo Galilei'' 
and CNISM, Universit\`a di Padova, \\
Via Marzolo 8, 35122 Padova, Italy \\ 
$^3$Dipartimento di Fisica, ETSF and CNISM,
Universit\`a di Milano, \\
Via Celoria 16, 20133 Milano, Italy} 

\begin{abstract} 
We show that both DC and AC Josephson effects 
with superfluid Fermi atoms in the BCS-BEC crossover 
can be described at zero temperature 
by a nonlinear Schr\"odinger equation (NLSE). 
By comparing our NLSE with mean-field extended BCS calculations, 
we find that the NLSE is reliable in the BEC side 
of the crossover up to the unitarity limit.  
The NLSE can be used for weakly-linked atomic superfluids 
also in the BCS side of the crossover by taking 
the tunneling energy as a phenomenological parameter. 
\end{abstract}

\maketitle

\section{Introduction} 

Few years ago the AC Josephson effect \cite{josephson,barone} 
with atomic superfluids 
was predicted \cite{smerzi} and observed \cite{jo-exp}
with Bose-Einstein Condensates (BECs). 
AC Josephson oscillations in superfluid atomic Fermi gases 
have been considered theoretically by several authors
\cite{paraoanu,wouters,adhikari,sala-jo}. 
Recently, Spuntarelli, Pieri and Strinati \cite{pieri-new} have studied
the DC Josephson effect \cite{josephson,barone} 
across the BCS-BEC crossover with neutral fermions by 
using the extended BCS equations: they computed the current-phase relation 
throughout the BCS-BEC crossover at zero temperature for 
a two-spin component Fermi gas in the presence of a barrier \cite{pieri-new}. 

In this paper we discuss a nonlinear Schr\"odinger equation (NLSE) 
\cite{sala-jo,manini05,sala-new} which is equivalent 
to the equations of superfluid hydrodynamics 
\cite{stringa-fermi} with the inclusion of a gradient term 
of the von Weizs\"acker type \cite{von,kirz,sala-kirz}. 
By using the NLSE we reproduce the DC Josephson 
results of Spuntarelli, Pieri and Strinati \cite{pieri-new}, 
in the BEC side of the BCS-BEC crossover, 
i.e.\ from the unitarity region to the deep BEC regime. 
However, we show that our NLSE describes \cite{sala-jo} both DC and AC 
Josephson effects \cite{josephson,barone} also in the BCS regime 
of two weakly-linked atomic superfluids 
if the tunneling energy is taken as a phenomenological 
parameter {\it \`a la} Feynmann \cite{feynman,packard}. 

\section{Hydrodynamics of Fermi superfluids at zero-temperature} 

At zero temperature the hydrodynamics equations of a 
two-component fermionic superfluid made of atoms of mass $m$ are given by 
\beqa 
{\partial \over \partial t} n &+& 
\nabla \cdot \left( n {\bf v} \right) = 0 \
\label{hy-1}
\\
m {\partial \over \partial t} {\bf v} &+& 
\nabla \left[ {1\over 2} m v^2 + U({\bf r}) 
+ { \mu(n,a_F)} \right] = 0  
\label{hy-2}
\eeqa 
where $n({\bf r},t)$ is the local density and ${\bf v}({\bf r},t)$ 
is the local superfluid velocity. 
Here $n({\bf r},t)=n_{\uparrow}({\bf r},t)+n_{\downarrow}({\bf r},t)$,  
with $n_{\uparrow}({\bf r},t)=n_{\downarrow}({\bf r},t)$ 
and ${\bf v}({\bf r},t)={\bf v}_{\uparrow}({\bf r},t)
={\bf v}_{\downarrow}({\bf r},t)$. 
$U({\bf r})$ is the external potential and 
$\mu(n,a_F)$ is the  bulk chemical potential, i.e. the 
zero-temperature equation of state of the uniform system, 
which depends on the fermion-fermion scattering length $a_F$. 
The density $n({\bf r},t)$ is such that 
\beq 
N = \int n({\bf r},t) \ d^3{\bf r} 
\eeq
is the total number of atoms in the fluid. In fact, due to the 
absence of the normal component, the superfluid density 
coincides with the total density and the superfluid current 
with the total current. 

Equations (\ref{hy-1}) and (\ref{hy-2}) are nothing but
the Euler equations of an inviscid and 
irrotational fluid. Since ${\bf v}$ is irrotational, 
it can be written as the gradient of a scalar field. 
The connection between superfluid hydrodynamics and 
quantum mechanics is made by the formula 
\beq  
{\bf v} = {\hbar \over 2m} \nabla \theta    \,,
\label{v-general} 
\eeq 
where $\theta({\bf r},t)$ is the phase of the condensate 
wave-function 
\beq 
\Xi({\bf r},t) = |\Xi({\bf r},t)| \ e^{i\theta({\bf r},t)} = 
\langle {\hat \psi}_{\uparrow}({\bf r},t) 
{\hat \psi}_{\downarrow}({\bf r},t) \rangle 
\,,
\eeq 
with ${\hat \psi}_{\sigma}({\bf r},t)$ the fermionic field operator 
with spin component $\sigma=\uparrow,\downarrow$ \cite{landau,leggett}. 
Notice the factor $2m$ (Cooper pairs) instead of $m$ in Eq.\ (\ref{v-general}) 
\cite{stringa-fermi,landau,leggett}.  

The equations of superfluid hydrodynamics describe quite accurately 
static properties and low-energy collective modes of oscillation
of wavelength $\lambda \gg \xi$,
where $\xi$ is the healing 
length of the superfluid. Recently, 
Combescot, Kagan and Stringari \cite{combescot} have suggested that
\beq 
\xi = {\hbar\over m v_{cr}} 
\label{healing}
\eeq
where $v_{cr}$ is the critical velocity of the Landau criterion 
for dissipation \cite{landau,combescot}. According to 
Combescot, Kagan and Stringari \cite{combescot},
in the BEC regime of bosonic dimers $v_{cr}$ coincides 
with the sound velocity, i.e. 
\beq 
v_{cr} = c_s = \sqrt{{n\over m}{\partial { \mu} 
\over \partial n}}  \; . 
\label{v-cr-b}
\eeq
Instead, in the BCS regime $v_{cr}$ is related 
to the breaking of Cooper pairs through the formula 
\beq 
v_{cr} = \sqrt{ \sqrt{{ \mu}^2 + 
|\Delta|^2}-{ \mu} \over m} 
\,,
\label{v-cr-f}
\eeq
where $|\Delta|$ is the energy gap of Cooper pairs \cite{combescot} 
Note that the equations of superfluid hydrodynamics 
(\ref{hy-1}) and (\ref{hy-2}) 
do not take into account the effect of pair breaking. 

In the BCS-BEC crossover the bulk chemical potential 
of Eq. (\ref{hy-2}) can be written as 
\beq 
{ \mu(n,a_F)} = {\hbar^2\over 2m} \left(3\pi^2 n\right)^{2/3} 
\left( { f(y)} - {y\over 5}{ f'(y)} \right) 
\label{mu-fermi} 
\eeq 
where $f(y)$ is a dimensionless universal function 
of the inverse interaction parameter 
\beq 
y={1\over k_F a_F} 
\eeq
where $k_F=(3\pi^2n)^{1/3}$ is the Fermi wavenumber \cite{manini05}. 
One can parametrize $f(y)$ as follows:  
\beq
{ f(y)} = \alpha_1 - \alpha_2
\arctan{\left( \alpha_3 \; y \;
{\beta_1 + |y| \over \beta_2 + |y|} \right)}  
\,,
\label{f-mc} 
\eeq
where the values of the parameters
$\alpha_1,\alpha_2,\alpha_3,\beta_1,\beta_2$, reported in Ref. 
\cite{manini05}, are fitting parameters based on asymptotics 
and fixed-node Monte-Carlo data \cite{giorgini}. 
We will call in the following Monte-Carlo equation of state (MC EOS)  
the equation $\mu=\mu(n,a_F)$ obtained from 
(\ref{mu-fermi}) and (\ref{f-mc}). 
 
Within the mean-field extended BCS theory \cite{leggett,marini}, 
the bulk chemical potential $\mu$ and the gap energy $\Delta$ 
of the uniform Fermi gas are instead found by solving the following 
extended BCS (EBCS) equations \cite{marini,sala-odlro} 
\beq 
-{1\over a_F} = {2 (2m)^{1/2} \over \pi \hbar^3} \,
\Delta^{1/2} \, 
\int_0^{\infty} dy \, y^2 \, 
\left(
{1\over y^2} - {1\over \sqrt{(y^2-{\mu\over \Delta})^2+1} }
\right)
\label{ebcs1} 
\eeq
\beq 
n = {N\over V} = {(2m)^{3/2} \over 2 \pi^2 \hbar^3} \,
\Delta^{3/2} \, 
\int_0^{\infty} dy \, y^2 \,
\left(
1 - {(y^2-{\mu\over \Delta})
\over \sqrt{(y^2-{\mu\over\Delta})^2+1} }
\right)
.
\label{ebcs2} 
\eeq 
By solving these two EBCS equations 
one obtains the chemical potential $\mu$ as a 
function of $n$ and $a_F$ in the full BCS-BEC crossover 
(see for instance Ref.~\cite{sala-odlro}). 
Note that EBCS theory does not predict the correct BEC limit: 
the molecules have scattering length $a_M=2a_F$ instead of 
$a_M=0.6a_F$ \cite{stringa-fermi}. We call EBCS equation of state 
(EBCS EOS) the mean-field equation of state $\mu=\mu(n,a_F)$ 
obtained from Eqs. (\ref{ebcs1}) and (\ref{ebcs2}). 
Of course, the MC EOS is much closer than the EBCS EOS 
to the MC results obtained in Ref. \cite{giorgini}. 

\section{Superfluid NLSE for the BCS-BEC crossover}

Inspired by the Ginzburg-Landau theory \cite{ginzburg}, by the 
density functional theory (DFT) \cite{dft}, and by the low-energy effective 
field theory (EFT) \cite{dicastro,son}, we introduce the complex wave function 
\beq
\Psi({\bf r},t) = \sqrt{n({\bf r},t)\over 2} 
\ e^{ i \theta({\bf r},t) } 
\label{psi} 
\eeq 
which describes boson-like Cooper pairs with the normalization 
\beq 
\int |\Psi({\bf r},t)|^2 d^3{\bf r} = {N\over 2} 
\eeq 
that is different from the normalization of the 
condensate wave function $\Xi({\bf r},t)$ \cite{sala-odlro}
while the phase $\theta({\bf r},t)$ is the same \cite{landau,leggett}. 
We now look for the simplest nonlinear Schr\"odinger 
equation of $\Psi({\bf r},t)$ which satisfies Eq. (\ref{v-general}) and  
reproduces the equations of superfluid hydrodynamics 
in the classical limit ($\hbar \to 0$). 
We find \cite{sala-jo,sala-new} that the nonlinear Schr\"odinger equation 
\beq 
i \hbar {\partial \over \partial t} \Psi({\bf r},t) 
= \left[ -{\hbar^2 \over 4 m} \nabla^2 + 
2 U({\bf r}) + 2 { \mu(n({\bf r},t),a_F)} \right] \Psi({\bf r},t) 
\label{super}
\eeq 
gives the equations of superfluid hydrodynamics  
in the classical limit ($\hbar \to 0$). 
For finite $\hbar$, this superfluid NLSE adds to the
classical hydrodynamics equations 
a quantum pressure term 
\beq 
T_{QP}=-{\hbar^2\over 8 m }
{\nabla^2 \sqrt{n}\over \sqrt{n}} 
\eeq
containing explicitly the  Planck constant $\hbar$ 
(gradient correction in DFT, next-to-leading correction in low-energy EFT) 
\cite{sala-new}. Note that in the deep BEC regime from Eq.~(\ref{super}) 
one recovers 
the familiar Gross-Pitaevskii equation for Bose-condensed 
dimers (molecules of two fermions), where 
\beq 
{ \mu(n,a_F)} = {4\pi \hbar^2 a_{dd}(a_F) \over 2 m} n 
\eeq
with $a_{dd}(a_F)$ the dimer-dimer scattering length, which 
depends on the fermion-fermion scattering length $a_F$. Within the 
mean-field theory one finds $a_{dd}(a_F)=2 a_F$, whereas 
four-body theory and also MC many-body data yield
$a_{dd}(a_F)=0.6 a_F$ \cite{stringa-fermi}. 

\section{Direct current Josephson effect} 

We use our time-dependent superfluid NLSE (\ref{super}) 
to study the direct-curent (DC) Josephson effect \cite{josephson,barone}. 
Consider a square-well barrier 
\beq 
U({\bf r})=\left\{ 
\ba{ll}
V_0 & \mbox{ for } |z|< d \\
0 & \mbox{ elsewhere } 
\ea \right. 
\label{barrier}
\eeq
which separates the superfluid into two regions, 
and assume a stationary solution 
\beq 
\Psi({\bf r},t) = \Phi({\bf r}) \ 
e^{i \theta({\bf r})} \ e^{-i2\bar{\mu}t/\hbar} 
\eeq
with constant and uniform number supercurrent
\beq 
J=n({\bf r}) {\bf v}({\bf r})=  
2 \Phi({\bf r})^2 {\hbar\over 2m} \nabla\theta({\bf r}) 
\,.
\eeq
From the previous equation it follows 
$(\nabla\theta)^2=m^2J^2/(\hbar^2\Phi^4)$ and also 
\beq 
\left[ -{\hbar^2 \over 4 m} \nabla^2 + 
{m\over 4}{J^2\over \Phi({\bf r})^4} + 
2 U({\bf r}) + 2 { \mu(n({\bf r}),a_F)} \right] \Phi({\bf r}) 
= 2\bar{\mu} \ \Phi({\bf r})
\,.
\eeq

\begin{figure}
\centerline{\psfig{file=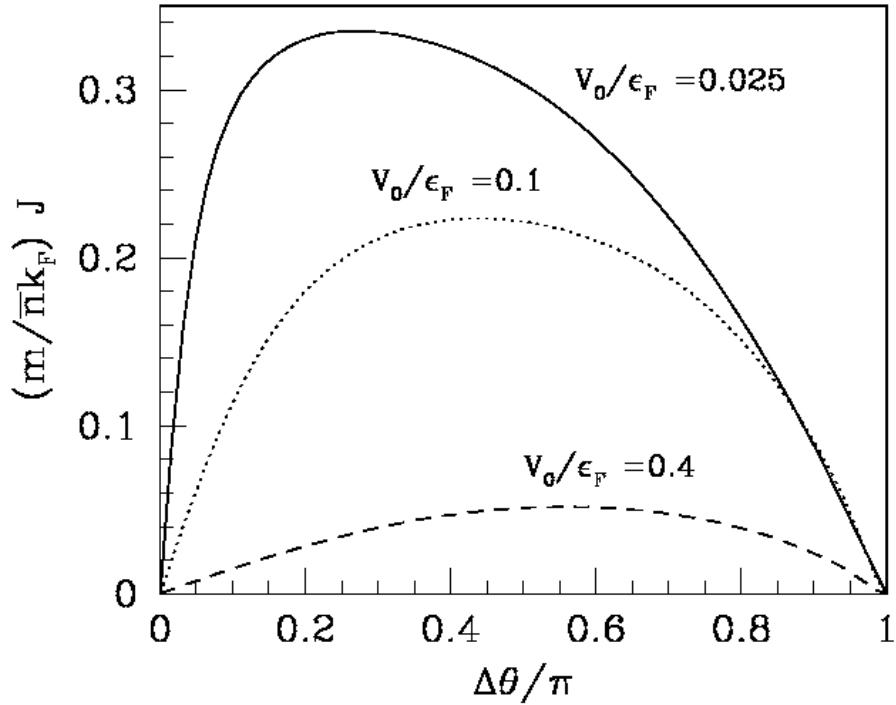,height=4.in}}   
\caption{DC Josephson current $J$ vs phase difference 
$\Delta \theta$ for a superfluid Fermi gas at 
unitarity ($y=0$), for three values of the energy barrier $V_0$. 
$\epsilon_F=\hbar^2(3\pi^2\bar{n})^{2/3}/(2m)$ is 
the Fermi energy. The width of the barrier is $L=4/k_F$, 
where $k_F=(3\pi^2n)^{1/3}$ is the Fermi wave number.} 
\end{figure}

We solve this stationary superfluid NLSE by imposing 
a constant and uniform density $\bar{n}$ at infinity:  
\beq 
\Phi({\bf r}) \to \sqrt{\bar{n}\over 2} 
\quad \mbox{for} \quad |{\bf r}|\to \infty 
\eeq
Given $\Phi({\bf r})$ at fixed $J$, 
the phase $\theta({\bf r})$ is then obtained from 
\beq 
\theta({\bf r}) = \theta({\bf r}_0) + 
{m J\over \hbar} \int_{{\bf r}_0}^{\bf r} 
{1\over \Phi({\bf r})^2} d{\bf r} 
\eeq
The phase difference across the barrier is defined as 
\beq 
\Delta \theta = \theta(z=+\infty) - \theta(z=-\infty) 
\,.
\eeq
This conditions allows us to establish the relationship between 
the current $J$ and the phase difference $\Delta \theta$. 
Figure~1 reports our results obtained by 
using the bulk EBCS EOS at unitarity ($a_F=\pm\infty$). 
As expected we recover the Josephson equation 
\beq 
J= J_0 \sin(\Delta \theta) 
\eeq
in the regime of high barrier (small tunneling, weak-link). 
In the limit of very small barrier (quasi-free transport, 
strong-link) $J_0$ has its maximum value, given by  
\beq 
J_0^{max}= \bar{n} \ v_{cr}   
\eeq
where $v_{cr}$ is the Landau critical velocity 
introduced in the previous section \cite{combescot}. 

\begin{figure}
\centerline{\psfig{file=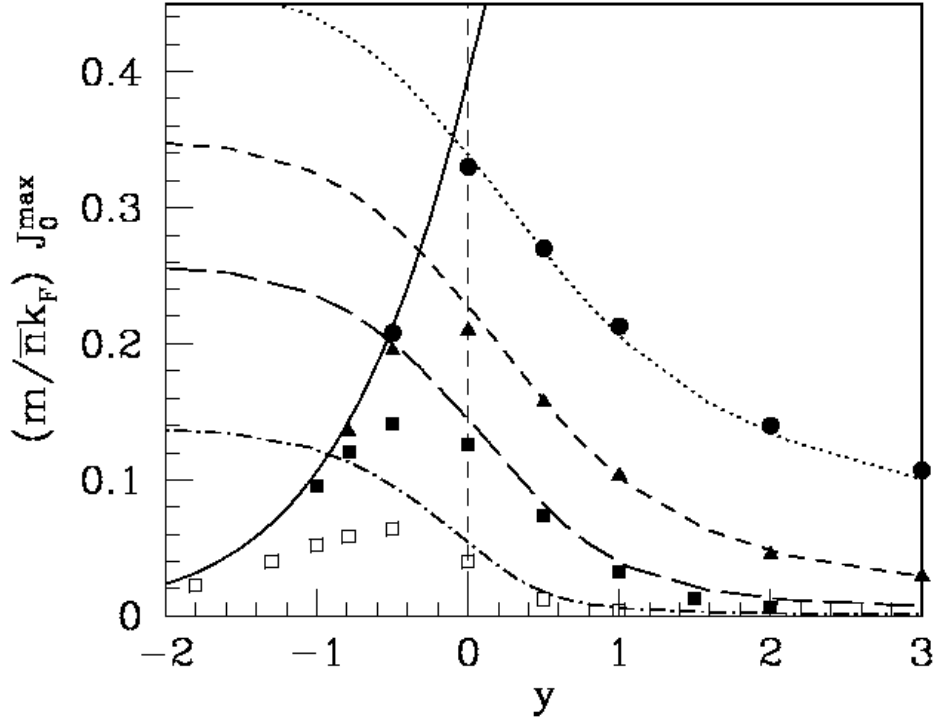,height=4.in}}                    
\caption{Maximum Josephson current $J_0^{max}$ vs inverse 
interaction parameter $y=1/(k_Fa_F)$ in the BCS-BEC crossover. 
Solid curve: $J_0^{max}$ based on pair breaking 
in the BCS regime \cite{combescot,pieri-new}. 
Other curves: superfluid NLSE. Symbols: microscopic mean-field 
calculations of Ref. \cite{pieri-new}. 
Four values of the energy barrier $V_0/\epsilon_F$ are considered: 
$0.025$, $0.10$, $0.2$, $0.4$. The width of the barrier is $L=4/k_F$.} 
\end{figure}

In Fig. 2 we plot $J_0^{max}$ as a function of the inverse interaction 
parameter $y=1/(k_Fa_F)$, and compare our data (curves) 
with the results of Spuntarelli, Pieri 
and Strinati (symbols) \cite{pieri-new}. 
Note that the data of Spuntarelli, Pieri and 
Strinati \cite{pieri-new} have been obtained by solving 
the full self-consistent Bogoliubov-de Gennes equations of the EBCS theory 
in the presence of the barrier (\ref{barrier}).  
Figure 2 shows that the NLSE reproduces the DC Josephson 
results of Ref. \cite{pieri-new}, 
but only in the right side of the BCS-BEC crossover, 
i.e. from the unitarity limit to the deep BEC regime. 
It is not surprising that the NLSE fails
in the BCS regime ($y<0$), since it 
neglects completely the effect of pair breaking.  
The critical velocity $v_{cr}$ predicted by NLSE 
is the sound velocity $c_s$ also in the BCS regime. 
Nevertheless, it is remarkable that the NLSE works 
quite well in the BEC regime ($y>0$) up to the 
unitarity limit ($y=0$). 

\section{Alternate current Josephson effect} 

We now consider a high barrier (small tunneling, weak-link),
without imposing a constant supercurrent $J$ and discuss 
the alternate-current (AC) Josephson effect \cite{josephson,barone} 
with superfluid Fermi atoms \cite{sala-jo}. 
We start from the time-dependent superfluid NLSE (\ref{super}) 
and look for a time-dependent solution of the form 
\beq
\Psi({\bf r},t) = c_A(t) \ \Phi_A({\bf r}) +
           c_B(t) \ \Phi_B({\bf r})  
\,,
\eeq 
where $\Phi_A({\bf r})$ and $\Phi_B({\bf r})$
is the quasi-stationary solutions normalized to one and 
localized in region $A$ and $B$ respectively. 
In this way we obtain\cite{sala-jo} the following two-state model 
\beqa
i\hbar {\partial \over \partial t} c_A(t) = E_A(t) \ c_A(t) 
+ { K} \ c_B(t)
\label{tun-a}
\\
i\hbar {\partial \over \partial t} c_B(t) = E_B(t) \ c_B(t) 
+ { K} \ c_A(t)
\label{tun-b}
\eeqa
for the two complex coefficients $c_A(t)$ and $c_B(t)$, related to
the number of atoms in the two regions. 
In our two-state model, $E_{A}(t)$ is the time-dependent 
energy in region $A$, 
given by  
\beq
E_A(t) \simeq 
\int \Phi_A({\bf r})
\left[-{\hbar^2\over 4m}\nabla^2 + 2U({\bf r}) + 
2{ \mu\left(2|c_A(t)|^2\Phi_A({\bf r})^2,a_F\right)} 
\right] \Phi_A({\bf r}) \, d^3{\bf r}
\,.
\eeq 
There is obviously a similar expression for the time-dependent 
energy $E_B(t)$. 
The constant coupling energy $K$ describes instead 
the tunneling between the two regions A and B: 
\beq
{ K} \simeq \int \Phi_A({\bf r}) 
\left[ -{\hbar^2\over 4m}\nabla^2 + 2U({\bf r})
\right]\Phi_B({\bf r}) \, d^3{\bf r}  
\,.
\eeq
 From our previous analysis of the DC Josephson effect, we expect 
that this expression is correct only in the right side 
of the BCS-BEC crossover. However, 
to extend the study the Josephson effect to the left side of 
the BCS-BEC crossover one may use $K$ in Eqs.~(\ref{tun-a}) 
and (\ref{tun-b}) as a phenomenological parameter 
{\it \`a la} Feynman \cite{feynman,packard}. 

\begin{figure}
\centerline{\psfig{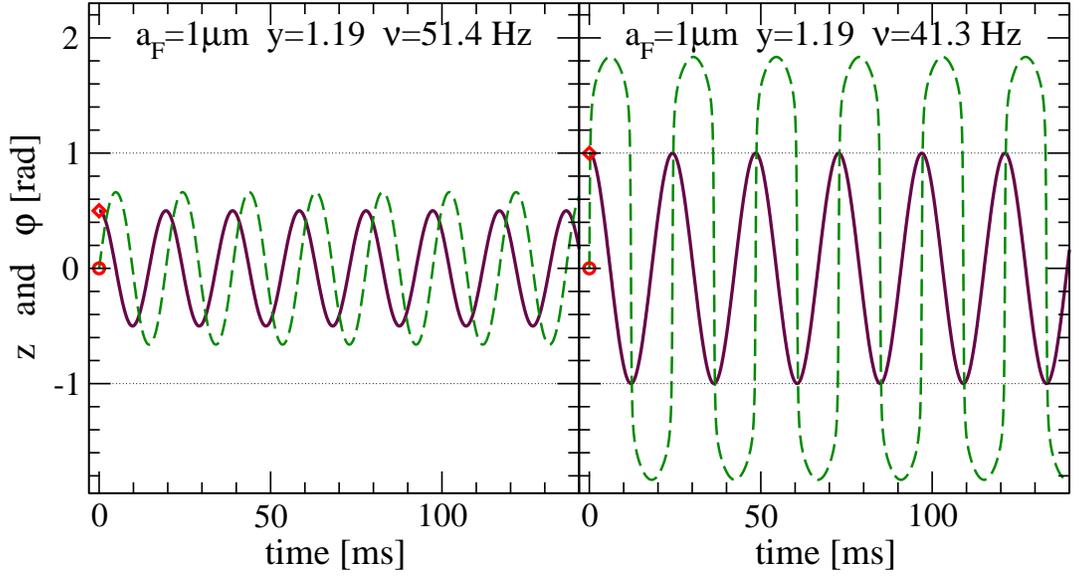}}  
\caption{Zero-mode in the AC Josephson effect by solving 
AJJ equations. $N=10^6$ $^{40}$K atoms between two symmetric regions 
of volume $25\cdot 10^6~\mu$m$^3$, 
tunneling parameter ${ K}/k_{\rm B} = 10^{-9}$\, Kelvin 
and Fermi-Fermi scattering length 
$a_F=1$~$\mu$m, corresponding to $y=1.19$. 
Solid line: population imbalance $z(t)$; dashed line: 
phase difference $\varphi(t)$. 
Initial conditions: $\varphi(0)=0$ and $z(0)=0.5$ (left); 
$\varphi(0)=0$ and $z(0)=0.999$ (right). 
Adapted from Ref. \cite{sala-jo}.} 
\end{figure}

We can write the complex coefficient $c_A(t)$ as 
\beq 
c_A(t) = \sqrt{N_A(t)\over 2} \ e^{i\theta_A(t)} 
\eeq
with $N_A(t)$ number of atoms 
and $\theta(t)$ phase in region $A$. Again, 
a similar expression holds for $c_B(t)$. In terms of the phase difference
\beq
\varphi(t) =\theta_B(t)-\theta_A(t)
\eeq
and relative number imbalance
\beq
z(t) ={N_B(t)-N_A(t) \over N_A(t)+N_B(t)} = {N_B(t)-N_A(t) \over N} 
\; ,
\eeq 
the two-mode equations give 
\beqa 
{\dot z}(t) &=& - {2{ K}\over \hbar} 
\, \sqrt{1 - z(t)^2} \, \sin\varphi(t) 
\label{eq-a}
\, ,
\\
{\dot \varphi}(t) &=& 
{2\over \hbar}\left[
{ \mu\left({N\over 2V}(1+z(t))\right)} -
{ \mu\left({N\over 2V}(1-z(t))\right)}
\right] + \! {2{ K}\over \hbar} \, {z(t) \over
\sqrt{1 - z(t)^2}} \, \cos\varphi(t)
\,.
\label{eq-b}
\eeqa 
These are the atomic Josephson junction (AJJ) equations 
describing the oscillations 
of $N$ Fermi atoms tunneling in the superfluid state between
region $A$ and region $B$, of equal volume $V$ \cite{sala-jo}.  
Notice that equations (\ref{eq-a}) and (\ref{eq-b})
generalize the BJJ equations
obtained by Smerzi {\it et al.} \cite{smerzi} 
for Bose-Einstein condensates. Moreover, 
these equations, linking the tunneling current 
\beq 
J = - {\dot z} {N\over 2} = 
{K N\over \hbar}\,\sqrt{1 - z^2}\,\sin\varphi =
J_0 \, \sqrt{1-z^2} \, \sin\varphi  
\eeq
to the phase difference $\varphi$, 
reduce to the familiar Josephson expression 
$J=J_0\sin(\varphi)$ in the appropriate limit $|z|\ll 1$. 

The nonlinear AJJ equations can be linearized around the 
stable stationary solution  
\beq 
\bar{z}=0 \quad \mbox{and}\quad {\bar \varphi} = 2 \pi j  
\eeq
where $j$ is an integer. In this way one finds the following frequency 
of small oscillation 
\beq
\nu_{0} = \frac {{ K}}{\pi \hbar} \sqrt{1 
+ {2mc_s^2 \over { K}}} 
\label{figata}
\eeq
which is called zero-mode. Here $c_s$ is the sound velocity 
computed at the mean density $n=N/V$ of the superfluid. 
This zero-mode is the analog of the Josephson plasma oscillation
in superconducting junctions \cite{barone}.
Figure 3 reports
the zero-mode oscillations of $z(t)$ and $\varphi(t)$, 
in the case of $^{40}$K atoms setting $a_F=1$ $\mu$m. 
The oscillation starting from $z(0)=0.5$ indicates that the solution
(\ref{figata}) of the linearized equations (\ref{eq-a}) and (\ref{eq-b})
are fairly accurate even for finite and not quite small amplitude.
Eventually however, for very large amplitude, $z(0)=0.999$, deviations
from the harmonic approximation become quite visible.

\begin{figure}
\centerline{\psfig{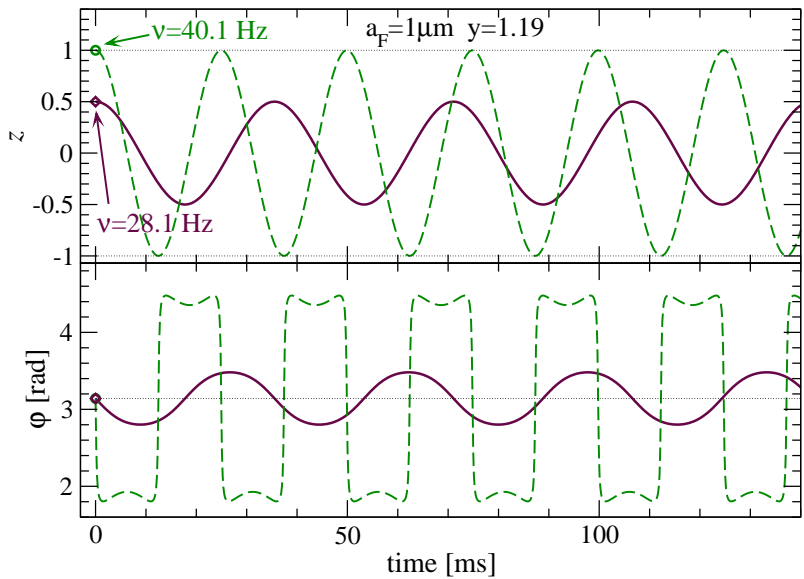}}    
\caption{$\pi$-mode in the AC Josephson effect by solving 
AJJ equations. All physical parameters are the same as in Fig.~3.
Upper panel: population imbalance $z(t)$; lower panel:
phase difference $\varphi(t)$. 
Initial conditions: $\varphi(0)=0$ and $z(0)=0.5$ (solid); 
$\varphi(0)=0$ and $z(0)=0.999$ (dashed).} 
\end{figure}

We observe that the AJJ equations (\ref{eq-a}) 
and (\ref{eq-b}) produce also a $\pi$-mode solution 
with $\bar{z}=0$ and $\bar{\varphi} =\pi (2j+1)$ and the self-trapping 
solution with population imbalance ($\bar{z}\neq 0$) (for details see 
\cite{sala-jo}).
Figure 4 reports $\pi$-mode oscillations for two different initial
unbalance.  Note the significant frequency increase and waveshape
distortion induced by nonlinear effects.

\section{Conclusions} 

This paper reviews a superfluid NLSE providing the hydrodynamic equations 
of Fermi superfluids plus a gradient correction. Both 
hydrodynamics equations and superfluid NLSE 
are reliable to investigate static properties and 
low-energy collective dynamics. It is important 
to stress that the equations of superfluid 
hydrodynamics are nothing else than the time-dependent local density 
approximation (LDA) for an irrotational system. 
The advantage of using the NLSE 
is that can take into account surface and shape effects beyond LDA,
and these can be relevant for a small number of particles \cite{sala-adh}. 
We have shown that in the study of the DC Josephson effect, 
the superfluid NLSE works 
quite well at the right side (BEC regime) of the BCS-BEC crossover up to the 
unitarity limit of infinite scattering length. 
In addition, we have suggested that 
for studying DC and AC Josephson effects in weakly-linked 
atomic superfluids the NLSE can be used in the full BCS-BEC crossover 
within the two-mode approximation.
In this case the tunneling energy coefficient must be taken as a
phenomenological parameter at the left side (BCS regime) of the crossover.

We thank Andrea Spuntarelli and Pierbiagio Pieri for making available 
their data \cite{pieri-new}. LS acknowledges a reseach grant from
GNFM-INdAM. This work has been partially supported by Fondazione CARIPARO.

\end{document}